\documentclass[conference]{IEEEtran}
\IEEEoverridecommandlockouts
\usepackage{cite}
\usepackage{amsmath,amssymb,amsfonts}
\usepackage{algorithmic}
\usepackage{graphicx}
\usepackage{textcomp}
\usepackage{xcolor}
\usepackage{comment}
\usepackage{hyperref}

\newcommand{\publicationNotice}{
    \begin{center}
        \fbox{
            \parbox{0.9\textwidth}{
                \textbf{Published Article:} This article has been published by IEEE in the Proceedings of 2022 48th Euromicro Conference on Software Engineering and Advanced Applications (SEAA), Gran Canaria, Spain, 2022, pp. 358-365. \textbf{DOI:} \href{https://doi.org/10.1109/SEAA56994.2022.00063}{https://doi.org/10.1109/SEAA56994.2022.00063}\\
                © 2022 IEEE.  Personal use of this material is permitted.  Permission from IEEE must be obtained for all other uses, in any current or future media, including reprinting/republishing this material for advertising or promotional purposes, creating new collective works, for resale or redistribution to servers or lists, or reuse of any copyrighted component of this work in other works.
            }
        }
    \end{center}
}

\newboolean{showcomments}
\setboolean{showcomments}{true} 
\ifthenelse{\boolean{showcomments}}
  {
		\newcommand{\nbb}[2]{
		\fcolorbox{black}{yellow}{\bfseries\sffamily\scriptsize#1}
		{\sf$\blacktriangleright$\textcolor{blue}{\textit{#2}}$\blacktriangleleft$}
		}
		
		\newcommand{\remarks}[1]{\color{red}[#1]\color{black}}

		\newcommand{\del}[1]{\textcolor{red}{\sout{#1}}} 
  }
  {
		\newcommand{\nbb}[2]{}
		\newcommand{\remarks}[1]{}

		\newcommand{\del}[1]{} 
  }

\def\BibTeX{{\rm B\kern-.05em{\sc i\kern-.025em b}\kern-.08em
    T\kern-.1667em\lower.7ex\hbox{E}\kern-.125emX}}

\begin{document}
\publicationNotice

\title{An Industrial Experience Report about Challenges from Continuous Monitoring, Improvement, and Deployment for Autonomous Driving Features
\thanks{Funded by Sweden's Innovation Agency, Diarienummer: 2021-02585}}

\author{\IEEEauthorblockN{Ali Nouri}
\IEEEauthorblockA{\textit{Volvo Cars} \\
Gothenburg, Sweden \\
ali.nouri@volvocars.com}
\and
\IEEEauthorblockN{Christian Berger}
\IEEEauthorblockA{\textit{University of Gothenburg, Sweden} \\
\textit{Department of Computer Science and Engineering}\\
christian.berger@gu.se}
\and
\IEEEauthorblockN{Fredrik Törner}
\IEEEauthorblockA{\textit{Volvo Cars} \\
Gothenburg, Sweden \\
fredrik.torner@volvocars.com}
}

\maketitle

\begin{abstract}
Using \emph{continuous development, deployment, and monitoring} (CDDM) to understand and improve applications in a customer's context is widely used for non-safety applications such as smartphone apps or web applications to enable rapid and innovative feature improvements.
Having demonstrated its potential in such domains, it may have the potential to also improve the software development for automotive functions as some OEMs described on a high level in their financial company communiqu\'{e}s. However, the application of a CDDM strategy also faces challenges from a process adherence and documentation perspective as required by safety-related products such as autonomous driving systems (ADS) and guided by industry standards such as ISO-26262 \cite{ISO26262} and ISO21448 \cite{sotif}.
There are publications on CDDM in safety-relevant contexts that focus on safety-critical functions on a rather generic level and thus, not specifically ADS or automotive, or that are concentrating only on software and hence, missing out the particular context of an automotive OEM: Well-established legacy processes and the need of their adaptations, and aspects originating from the role of being a system integrator for software/software, hardware/hardware, and hardware/software. In this paper, particular challenges from the automotive domain to better adopt CDDM are identified and discussed to shed light on research gaps to enhance CDDM, especially for the software development of safe ADS. The challenges are identified from today's industrial well-established ways of working by conducting interviews with domain experts and complemented by a literature study.
\end{abstract}


\section{Introduction}
The number of collisions for vehicles equipped with advanced driver assistance systems (ADAS) led to a reduction in serious injuries and deaths \cite{magdacitysafety,magdaVRU,who}. ADAS support the driver to avoid collisions by providing warnings, automatic steering, and braking interventions but the driver is still supervising the functions \cite{SAE} and is legally responsible for the entire driving task. 
However, this trend seems to plateau and hence, demanding now the next generation of safety systems to further reduce the risk of such accidents to happen \cite{who}. 
Autonomously driving vehicles (AD) that remove the human driver from the driving task are a prosperous path towards achieving Vision Zero \cite{visionzero}, ie., no fatal or severe road accidents in the future. However, replacing a human driver with an unsupervised intelligent system increases at the same time the complexity of the overall system drastically, implying higher risks of safety issues \cite{ric_av_safety}. The complex nature of the operational design domain (ODD) for AD includes various road users such as other vehicles or pedestrians and bicyclists. This increased complexity of the problem space challenges and the argumentation that an AD function is \emph{designed safe and remains safe}.

\subsection{Background}
\begin{figure*}
  \includegraphics[width=\textwidth]{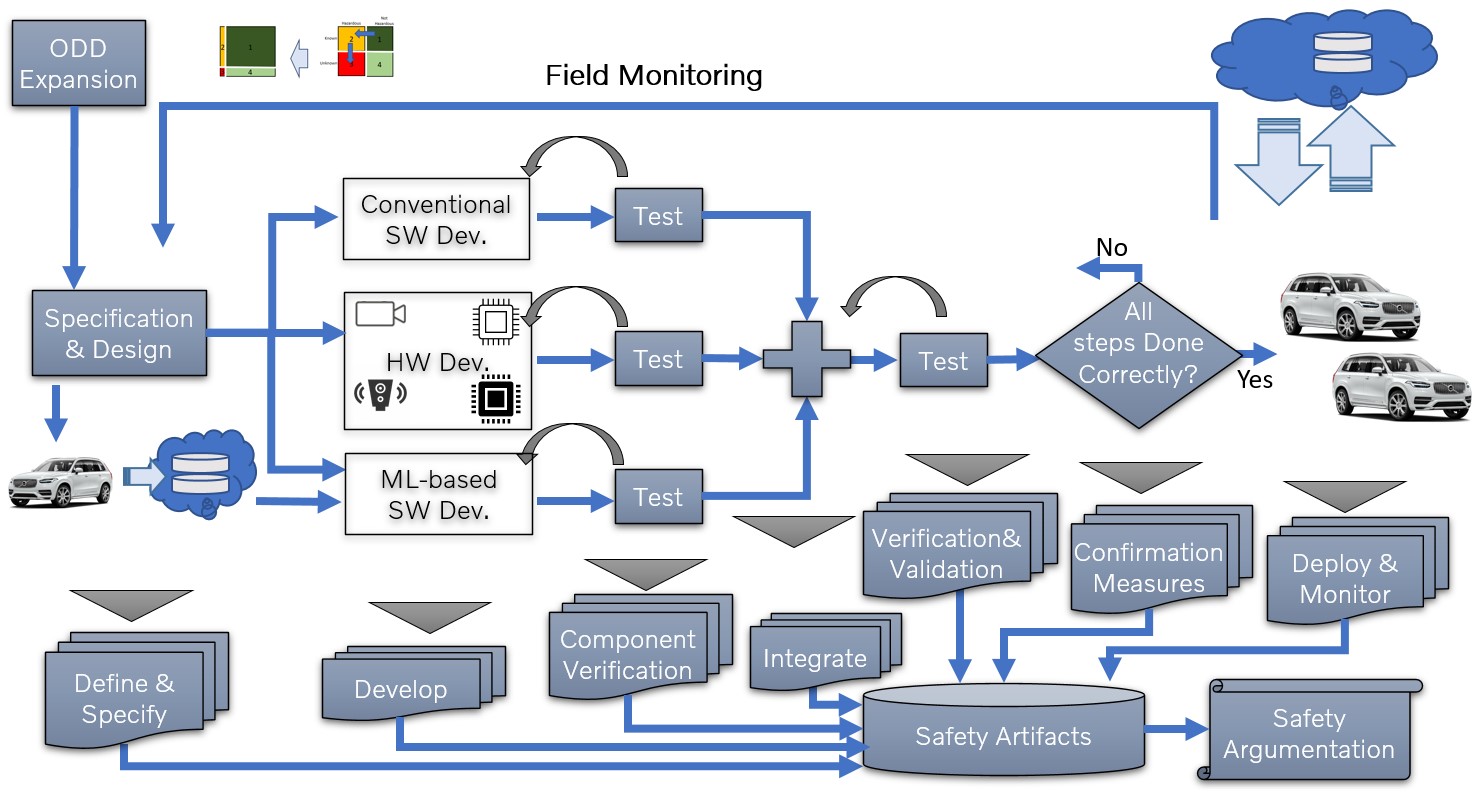}
  \caption{Iterative development process for autonomous driving and its contribution to safety argumentation.}
  \label{fig:Overview}
\end{figure*}

\begin{figure*}
  \includegraphics[width=\textwidth]{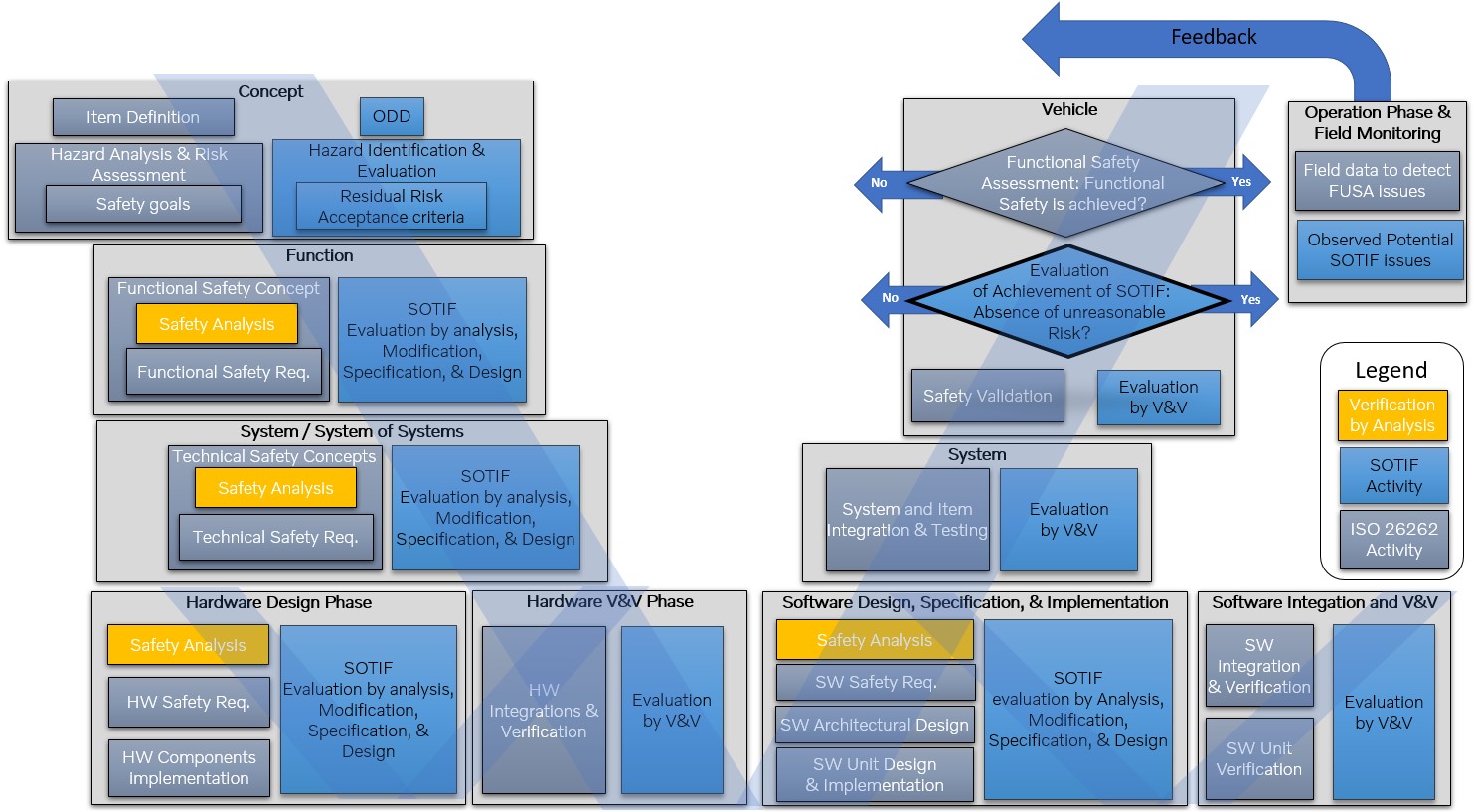}
  \caption{ISO 26262 and SOTIF Activities in each abstraction level}
  \label{fig:FUSASOTIF}
\end{figure*}

The development of a safety-relevant function begins with a traceable requirements specification at different abstraction levels (eg., safety goal, functional safety requirements, technical safety requirement, and hardware/software requirements)\cite{ISO26262} in parallel with performing different kinds of analyses like Fault Tree Analysis (FTA), Failure Mode and Effect Analysis (FMEA), Systems Theoretic Process Analysis (STPA) to verify the safety of solutions before their implementation as depicted in Fig.~\ref{fig:Overview}. Then, the requirements allocated to each of these elements (ie., system or hardware/software component) will be implemented and integrated with each other in various abstraction levels from component up to the complete vehicle. The next step is to verify each safety requirement to ensure that the system's implementation and integration has been conducted according to the specification, followed by validation steps to ensure that requirements are complete and correct. Finally, all these steps must be reviewed and assessed by a sufficiently independent assessor before an official release. Fig.~\ref{fig:FUSASOTIF} depicts the aforementioned process within ISO 26262 and SOTIF for the various abstraction levels.

\subsection{Problem Domain \& Motivation}
Even if all aforementioned steps are strictly followed, an automated driving (AD) system will surely be exposed to various safety-related hazards during its operational phase \cite{isotr4804}, which shed light on the need for continuous monitoring of a function once deployed to customers. Hence, the development of the actual ADS is now stretching into its operational phase after the cars left the factory to better cope with challenges originating from the sheer unbounded complexity of scenarios, components, and their integration to make \emph{and keep} an ADS safe. This motivates continuous monitoring and adaptation of an already deployed function in its ODD using agile ways of working in combination of DevOps to enable a gradual and safe function growth and ODD expansion.

The \emph{quality monitoring and gradual feature improvement/expansion} way of working have been used successfully in non-safety-related domains. Prominent examples such as Google, Facebook, and Spotify utilize countless data points from observing their customers' behavior when they are interacting with their products. The approach to continuously monitor and improve a product enables software developers to systematically experiment with new ideas on a smaller group of users to observe key performance indicators (KPIs) and adoption rates before rolling out a new feature to all customers. While this approach, at first sight, seems like a ``trial and error'', it is already the key differentiating way of working to accelerate a company's level of innovation and to stay ahead of competitors as indicated by annual DORA reports.\footnote{Cf.~Google: ``What is DevOps?'', \url{https://cloud.google.com/devops}} 
However, the aforementioned example target non-safety-critical applications, and hence, the motivation is on what aspects therefrom can be carried over to safety-critical system development with its inherent requirements on established integrity and safety.

CDDM is not unknown and was conducted even for safety-related hardware technologies and data therefrom was later used for the next improvement in the product. The only difference is the speed of these iterations, which changed from months to potentially days. However, before implementing any change in the function or system, an impact analysis needs to be performed. The impacts of any change will be identified and as a result, the safety activities must be tailored accordingly.

\subsection{Research Goal \& Research Questions}
The aim of this research is to identify and report about challenges reported by industrial practitioners related to field monitoring, continuous development, and deployment of AD applications under safety considerations. We break down our research goal into the following research questions:

\begin{description}
  \item[RQ-1] What are safety-related challenges during field monitoring of AD?
  \item[RQ-2] What are safety-related challenges for continuous development and deployment?

\end{description}

\subsection{Contributions}
In this work, we are reporting the challenges originating from the safety perspective as perceived by industrial practitioners addressing continuous monitoring, improvement, and deployment of autonomous driving. We have digested these insights from stakeholder interviews with automotive experts with years-long experience that we enriched and complemented with aspects from standards, and regulatory documents.

\subsection{Scope}
This study is focusing on the context of Functional Safety, SOTIF, and safety aspects of cyber-security (CS) for the rapid development of safety critical AD functions. While selected challenges and ideas may be applicable to other contexts as well, we are only focusing on the automotive domain primarily from an OEM's perspective. While we are starting with vehicle level challenges, we assume that electrical/electronic hardware aspects cannot be updated during an AD life-time, and hence, we scope it primarily down to software-related aspects.

\subsection{Structure of the Article}
The remaining sections of this article are structured as follows: In Sec.~\ref{sec:background}, fundamental and necessary background definitions and terms are introduced and put into context. We discuss related work in Sec.~\ref{sec:relatedwork} and report about our methodology in Sec.~\ref{sec:methodology}. We describe and discuss the identified challenges in Sec.~\ref{sec:results_discussion} before we conclude our work in Sec.~\ref{sec:conlusion} and provide an outlook for future work.


\section{Background}
\label{sec:background}

The overall achieved safety of an AD function can be understood from various perspectives: 
\underline{Functional Safety}: Absence of unreasonable risks by avoiding malfunctions caused by systematic failures in hardware and/or software, or random hardware faults in the implementation \cite{ISO26262}. The ISO-26262 provides objectives and requirements supported by recommendations in order to enable a user to argue for achieving the absence of unreasonable risk. 

\underline{Safety of intended function (SOTIF)}\cite{sotif}:  
As it is shown in Fig.~\ref{fig:FUSASOTIF} SOTIF asks for determining acceptance criteria which is represented by validation target regarding insufficient specification or performance limitations. Then during the ``evaluation by verification and validation'' phase, the absence of unreasonable risks will be evaluated. SOTIF is categorizing scenarios into the following four groups: \emph{Hazardous/Known}, \emph{Hazardous/Unknown}, \emph{Not Hazardous/Known}, and \emph{Not Hazardous/Unknown}, and the goal is to reduce the exposure of scenarios from the two groups \emph{Hazardous/Known} and \emph{Hazardous/Unknown} to an acceptable level of risks before a function is released. SOTIF requires or recommends activities but yet, some level of risk may remain due to changes in assumptions during the design phase or unidentified triggering conditions. According to SOTIF, the field monitoring process shall be in place before the release of a function and would be used to maintain the safety of the intended functionality \cite{sotif}.

\underline{Cyber-Security}: In this study, only the unsafe effect of malicious intent from an external source to the system is addressed \cite{ISO26262}. Since cyber-security issues and also sometimes their solutions like bug-fixes can violate the safety requirements, ISO-26262 is asking for interactions between the management of functional safety and cyber-security during the field monitoring phase for reported incidents. The interaction between these contexts is not limited to field monitoring but it also includes activities from the design phase as hazards and threats need to be cross-checked to make sure they are considered in both. 

\underline{Over-the-air} (OTA): Continuous connectivity allows not only monitoring and diagnostics but also enables faster software updates after the vehicles have left the factory. Companies integrate all necessary hardware to the vehicle without considering the current software being ``final'' but rather that further functionality will be available when reaching a certain maturity level that is measured systematically by using field data. When the software is deployed and the user can activate a function, data is collected and where legally required anonymized so that the OEM can monitor the performance, safety, and other aspects of the function. Therewith, the safety of the software will be maintained, and the needed evidence for arguing the safety of the next functions release is gathered. Examples from the industry showed that some level of AD was implemented and due to unsafe behavior of the intended functionality, the manufacturer performed a recall and updated the fixed software by means of OTA \cite{tesla22}.


\section{Related Work}
\label{sec:relatedwork}
Safety of AD has become a growing challenge during the last decade, especially after tragic accidents during public testing that made research groups and industrial organizations to increase their activities to significantly improve the safety during the development and operations of such functions.
Zeller started to shed light on challenges for safety-critical applications regarding DevOps way of working \cite{zellersafedevops}, but it is yet immature in the automotive domain due to complex product structures \cite{Rolf2019Cont}.
The Swedish research project SALIENCE 4CAV\footnote{SALIENCE4CAV project: https://salience4cav.se/} is aiming at finding solutions for the safety life-cycle of AD to enable continuous deployment . AI and ML as the main reasons for CDDM attracted significant attention like KI-Absicherung (German for ``assuring AI'')\footnote{KI-Absicherung: https://www.ki-absicherung-projekt.de/en/project}, which is dealing with the safety life-cycle of ML in the context of AD. Willer et al.~identified nine safety concerns like ``distributional shift over time'', and ``unknown behavior in rare critical situations'' \cite{willers2020safety}, where relevant mitigation approaches are gathered and linked to safety concerns. Although all proposed mitigation strategies are applicable also in an iterative development, ``continuous learning and updating'' and ``deep analysis of test results obtained in an iterative development process'' are only applicable in an iterative approach. S\"{a}mann et al.~also emphasized the necessity of CDDM like proposing online monitoring during run-time to tackle uncertainties and out-of-distribution challenges \cite{samann2020strategy}.

The currently ongoing SMILE program\footnote{SMILE: https://www.ri.se/en/what-we-do/projects/smile-iii-safety-analysis-and-verificationvalidation-of-ml-based-systems} is researching methods to assure the safety of ML-based software to be able to use them in safety-critical functions. This is relevant work from an industrial perspective, as many OEMs and tech companies have demonstrated prototypical vehicles with AD features, while younger start-ups do not have the challenges from legacy software stacks and are known for speed and flexibility in product enhancements\cite{knauss2016continuous}.

There are mature methods such as those prescribed in ISO-26262 but they are not enough regarding arguing for the safety of complex systems that include AI/ML elements. This is why there is ongoing industrial standardization to define sufficient methods to create relevant processes like safety for automated driving systems (ISO TS 5083)\footnote{Under development. Stage at the time of publication: ISO/AWI TS 5083. https://www.iso.org/standard/81920.html}, Safety Of Intended Functionality (SOTIF), and Safety and Artificial Intelligence (ISO PAS 8800)\footnote{Under development. Stage at the time of publication: ISO/AWI PAS 8800. https://www.iso.org/standard/83303.html}.

But yet, unique technical challenges and relevant solutions for designing, implementing, verifying, and validating a safe AD through continuous field monitoring are under-explored. Regulators also closely monitor this field and dynamically release regulations for the safety of AD functions, which OEMs are liable for in each market, to assure that best practices are followed. For example, to release a feature such as ``Automated Lane Keeping System'' (ALKS)\cite{ALKS} for the European market, the auditor and assessor of the product shall be competent in standards such as ISO-26262 \cite{ISO26262}, ISO-21448 \cite{sotif}, and ISO-21434 \cite{isocs}. 


\section{Methodology}
\label{sec:methodology}
In this industrial experience report, we are combining (A) a sequence of interviews with domain experts from industry and (B) findings of the current state of practice reported in literature and regulations that are both, well established or currently under development.

The interviews were planned as follows: Firstly, questions about the problem domain and potential attempts to overcome challenges therefrom were identified and refined within the research team resulting in the following structure of questions that were used in open-ended interviews with industrial practitioners:

\begin{enumerate}
    \item What is your role in the company?
    \item How long are you active in that role?
    \item What are the main challenges of CDDM in the AD context?
    \item Please comment on the general challenges (cf.~list in Sec.~\ref{sec:results_discussion}).

\end{enumerate}

The interviews were planned for approximately one hour with the possibility for extension and conducted both, on-site and online. We identified 8 interviewees by convenience sampling from our industrial network and during community gatherings of automotive safety experts. Anonymization and a round of clarification with the notes were assured to the interviewees. During the interviews, central statements were noted down and confirmed subsequently with the interviewees to capture the intended ideas and opinions. Using this strategy prevented the need to apply clustering approaches like \emph{word clouds} as key ideas were already identified during the interviews and confirmed afterward. The findings from these interviews were then used to complement, extend, or contradict what was identified in regulations and literature.

The second part of our method consisted of a literature review of regulations relevant to industrial practitioners and related literature thereto. The documents and papers were identified as described in the following and key recommendations were extracted to establish the methodological setup recommended to industrial practitioners:

\begin{itemize}
    \item Regulations, working drafts for regulations for AD and commentaries thereto: Standards (ISO 26262, ISO21448, ISO TS 5083, ISO PAS 8800), and further related publications (eg., SaFAD, safety reports, competitors analysis)
    \item Involvement in discussions from standardization efforts (ISO TS 5083, and ISO 8800) allowed access to publications and further expert opinions
    \item Stakeholder interviews with industrial experts and practitioners
    \item Performing snow-balling on selected relevant publications

\end{itemize}

Finally, we combined the extracted key statements and ideas from the interviews and essential recommendations from regulations and documents for industrial practitioners to identify potential methodological issues or missing methodological recommendations to derive research gaps and potential research directions. The challenges and the effectiveness of already existing approaches from an OEM's perspective to deal with CDDM in the AD domain are extracted. In order to identify the challenges, the automotive engineering process is analysed and potential process inhibitors at the vehicle level (e.g., requirements specification, vehicle integration, and validation) to software and hardware components (e.g., implementation, dimensioning requirements, verification) are identified as depicted in Fig.~\ref{fig:FUSASOTIF}. We report and discuss the insights and results from this extraction in Sec.~\ref{sec:results_discussion}.


\section{Results \& Discussion}
\label{sec:results_discussion}

In the following, we present the aggregation of our insights from the interviews with the technical experts from the field and complement them with key statements from regulations or commentaries thereto. In total, we interviewed two technical experts, a solution architect, a product owner, a product manager, a software process developer, a verification engineer, and a function developer.

\begin{tabular}{|p{0.9\columnwidth}|}
  \hline
  \textbf{CH1: Impact analysis and tailoring before each iteration} \\  
  \hline
\end{tabular}

As part of the planning for any change in function, system, or software abstraction level, an impact analysis shall be performed to identify the effects of the change on elements involved in providing the function. Then, based on a rationale, some activities in each iteration can be omitted or performed in a different way, which in the ISO-26262 terminology is named tailoring. Tailoring and its rationale will be used in the safety argumentation (ISO 26262 part 2 clause 6) \cite{ISO26262}, the rest of the activities needed to be done again, and the relevant work products shall be updated.

\emph{Discussion:} Industrial experience showed that such impact analyses on functional levels are time-consuming and resource intense. Hence, the industry may tend to gather multiple changes in one batch before conducting another impact analysis (IA). While doing so may reduce the actual effort to avoid doing IAs too frequently, the actual gathering of functional changes is delaying their roll-out and hence, valuable data-points from the ODD are only available later to the development teams who, though, need insights from the field.


\begin{tabular}{|p{0.9\columnwidth}|}
  \hline
  \textbf{CH2: Requirement updates (FUSA, SOTIF, or Cyber-Security)} \\  
  \hline
\end{tabular}

In the context of ISO-26262 after defining an item, the first step is Hazard Analysis Risk Assessment (HARA), which will provide the first level of safety requirements: The safety goals (SGs) \cite{ISO26262}. These SGs will then be broken down in the next abstraction levels and finally be allocated to hardware components or software units to be implemented. For SOTIF, also the item definition will be the input, which leads to defining the validation targets (VT) and acceptance criteria. To address cyber-security, the Threat Analysis, and Risk Assessment (TARA) derives the cyber-security goals. If the definition of an item is changed in an iteration (eg., function or ODD expansion), each analysis shall be checked by impact analysis, to identify the need for any change. Moreover, new hazards or threats might be identified during the operation phase, which were unknown before and hence, not considered so far. They need to be added to the relevant catalog and analyzed resulting in the need to add new safety goals, validation targets, or cyber-security goals.

\emph{Discussion:} Conducting a HARA is time-consuming and requires several weeks to months in practice. For example, to adjust slightly the addressed speed range would require a revised (or even a new) HARA. Hence, such time-consuming and resource-intense analysis needs to be planned ahead and would not easily allow for an agile and rapid refinement based on insights from field data obtained from remote system monitoring.


\begin{tabular}{|p{0.9\columnwidth}|}
  \hline
  \textbf{CH3: Change in one discipline affects others} \\  
  \hline
\end{tabular}

Any change in one of the disciplines (e.g., FUSA) has effects on the others (e.g., CS), which is needed to be considered. In some cases, the safety design (both FUSA and SOTIF) and cyber-security design are competing, and hence, there is a need to consider all these three disciplines simultaneously\cite{ISO26262} and this is exactly what a developer is doing when performing system design (i.e. finding one solution for all requirements, not only FUSA/CS/SOTIF but also cost and availability for example).

\emph{Discussion:} Considering the needed time and effort to perform impact analyses and update the requirements accordingly would reduce the speed of each iteration. Therefore, not only the top-level analyses in FUSA (ie., the HARA), SOTIF (ie., the HIdEv), and cyber-security (ie., the TARA) shall be aligned, but also other design activities on different abstraction levels (e.g., FSC, TSCs, and CS Concepts) to avoid mismatch or competition between solutions in the individual contexts. For example, a security mechanism can violate a safety requirement or vice versa, and hence, identifying such competing needs is again time-consuming and resource-intense.


\begin{tabular}{|p{0.9\columnwidth}|}
  \hline
  \textbf{CH4: Hardware limitations} \\  
  \hline
\end{tabular}

Hardware components of systems involved in providing AD functionality like sensors, processors, and actuators are difficult to upgrade or not upgradeable at all without having the customer to come to a workshop. For non-safety-critical applications, the function might be delivered partially (eg., by using feature toggles and reducing risks from delayed integrations) even if the hardware is not fully capable of performing the task. But for safety-relevant functions, this must be avoided since it may lead to a hazard. By upgrading the function, the requirements may change and their ASIL may increase as a consequence, and hence, the hardware must satisfy the new ASILs and, if applicable, ASIL decomposition requirements may be added. For example:
\begin{itemize} 
    \item The sensor set might not be capable of performing perception properly for the next generation of the function: If the speed limit is increased the sensor set needs to cover longer distances or the sensor set does not have enough diversity and redundancy to satisfy new ASIL decomposition pattern. 
    \item The computing unit needs to be capable of handling new versions of software architecture and thus, sufficient processing computational power must be available.
\end{itemize}

\emph{Discussion:} Besides the upgradability constraints of the hardware during a vehicle's life-cycle, the timelines during the vehicle development may not be perfectly aligned between hardware and software development resulting in further coordination activities that may impact the analysis and integration steps. Furthermore, ODD expansions from practice as the OEMs see the upcoming potential for providing more features to customers is challenging a previously agreed hardware and system design, which may lead to missed opportunities for innovations.

\begin{tabular}{|p{0.9\columnwidth}|}
  \hline
  \textbf{CH5: Safety analysis methods} \\  
  \hline
\end{tabular}

When a design in any abstraction level changed, there will be a need to analyse the impact of the design change on performed analyses such as FMEA, FTA, or STPA. Based on the outcome some might need to be updated based on the new design. As an outcome of the update new requirements and safety mechanisms might be added.

\emph{Discussion:} The growing complexity of such a software-intense system also increases the difficulty of conducting these analyses, which, in turn, require more time and resources. So, there is a direct dependency between the growing amount of ``expected'' functionality from a customer's perspective and the effort during development and operation: While the first aspect is needed to remain competitive in the market, there are clear needs to find more agile approaches during the development and especially during the monitoring and operations phase of a system, without compromising on overall system safety.


\begin{tabular}{|p{0.9\columnwidth}|}
  \hline
  \textbf{CH6: Software architecture} \\ 
  \hline
\end{tabular}

The design of the software architecture is among the first steps during the software development. The software architecture shall contain the software components and the interaction between them even if they are allocated to different ECUs or processors\cite{ISO26262}. Any change in the software interfaces or interactions between software elements will affect the software architecture and will lead to the need of redoing the relevant activities like safety analysis, requirements specifications, verification, and validation. In some cases, a change in the software architecture may have an effect on the software allocation to hardware, which leads to the need for additional effort in order to perform safety activities on the system and hardware.

\emph{Discussion:} The growing complexity of software requires a paradigm-shift in the way how software architectures are created and treated during the development: While activities such as AUTOSAR in the past focused on clear separations of concerns to abstract as much as possible hardware and networking from software units, today's increased computational needs require a rethinking of overcoming the software isolation from its hardware platform. Recent initiatives such as SOAFEE\footnote{\url{www.soafee.io}} aim at addressing the orchestration of software to hardware capabilities to better meet the increased computational needs and hence, the software architecture needs to inevitably become more aware of the hardware resources and features, as well as need to become more flexible than rigid while adhering to the expected safety goals.


\begin{tabular}{|p{0.9\columnwidth}|}
  \hline
  \textbf{CH7: Verification} \\ 
  \hline
\end{tabular}

The verification is needed to assess if the implemented element meets its requirements. Verification can be done by means of review, analysis, simulation, or testing \cite{ISO26262}. During each iteration, the affected function, system, or software elements need to be verified against the requirements after having implemented the software in case of refining the safety requirements on any abstraction level. Before deploying a new software element, the manufacturer shall verify that the new element does not affect any other unchanged elements (ie., regression testing).

\emph{Discussion:} This is the main challenge as mentioned in SOTIF and also well known in the industry resulting in redoing verification and validation activities. Especially for AI/ML-enabled features whose engineering is driven by data, its labeling, and training/re-training, the right combination of tools for monitoring and performance evaluation during the development \emph{and} for any change during its operation in the field is still an enormous challenge for industry. Connected thereto is the quest to strive for explainability of AI/ML-systems that is a continuously persisting challenge, which both, academia and industry are currently trying to address. The particular challenges here are scenario identification for reproducibility of cause-effect-analysis, as well as the quantification of fidelity gaps between virtual testing approaches such as simulations and expected behavior, in reality, using the production hardware setup.


\begin{tabular}{|p{0.9\columnwidth}|}
  \hline
  \textbf{CH8: Validation} \\ 
  \hline
\end{tabular}

Validation in the FUSA context is to assure that the safety goals are adequate and achieved. In SOTIF, it will be used to assure that the product is free from unreasonable risks defined by qualitative or quantitative acceptance criteria. When a function or its ODD is expanded, there might be a need for (re-)validating the new feature, or a new version of the function is required entirely, which is one of the biggest challenges in AD (eg., Hazardous/Unknown scenarios). 

\emph{Discussion:} The challenge for the automotive context is here to balance opportunity identification from field monitoring to expand on functionality or ODD, while reducing the (re-)validation efforts. Compared with the previous challenge, similar aspects need to be considered in the industry.


\begin{tabular}{|p{0.9\columnwidth}|}
  \hline
  \textbf{CH9: Safety argumentation} \\ 
  \hline
\end{tabular}

Argumentation consists of three main groups of elements: Claim, evidences, and strategy, which connects the evidences to the claim in a way to prove that the claim holds. In our context, the claim is a set of safety goals or validation targets, and evidences are the safety artifacts. Through different strategies, the top-level goal will be broken down into sub-goals (eg., software or hardware safety requirements), which in the end will be supported by a set of evidences. During each iteration, both safety artifacts and top-level safety objectives will be updated, which might lead to the need of updating the strategies and safety argumentation itself. The safety argumentation shall be understandable, maintainable, and expandable/reusable for future applications \cite{kelly1999arguing}. On the other hand, using a natural language for presenting the strategies in a safety argumentation by using, for instance, narrative text is a common approach, but challenging in order to maintain it during each iteration. Moreover, most of the safety artifacts are spread in different tools in practice, which makes it time-consuming and hard to keep the traceability to updated artifacts \cite{zellersafedevops}.

\emph{Discussion:} This is particularly challenging in practice as besides being time-consuming and resource-intense, country-dependent regulations also needs to be considered that may disallow certain type of system modifications when type-approvals are needed. Furthermore, industrial practitioners pointed out that safety requirement traceability, while uttermost important for the safety argument, is still challenging in practice. This means, though, that a proper artifact management between the various departments in an organization as well as covering the various development and operation phases for a vehicle is needed.

\subsection{Concluding Comments}
As shown by the presented challenges, they are spanning all aspects of product development, including impact analysis, design, analysis, verification \& validation, and the safety argumentation, ie., product documentation. The main commonality in the challenges is the involvement of intellectual work that is required for every proposed or intended change of the product. Hence, an overall conclusion about the challenges of introducing CDDM for safety-relevant products is to explore ways how to support the human in doing these intellectual activities and to utilize automation where possible to meet rapidity expectation from competitive perspectives.
\subsection{Threats to Validity}
\subsubsection{Construction Validity}
As outlined in Sec.~\ref{sec:methodology}, the key statements and ideas were identified already during the interviews and confirmed afterwards with the interviewees instead of clustering using interview transcriptions and word frequency analysis afterwards. The risk, though, is that some key ideas may be missed as a verbatim transcript would preserve the flow of an interview. Furthermore, spotting such key ideas require domain familiarity of the interviewer. We consider, though, this being the case in our research context. In addition, any minor misunderstandings or aspects that would need further clarification were corrected by presenting the key Challenges or statements with the interviewees afterwards. 

\subsubsection{Internal Validity}
The study was designed as an experience report and hence, the insights, findings, and conclusion undeniably depend on the sampling of the interviewees. Furthermore, all involved authors have years-long experience from working in or with the automotive industry as well as working with safety-related functionality, which may result in potentially overlooking obvious aspects. We have tried to address this by finding a balance between the academic and industrial involvement in the research and methodological design, as well as in the interpretation of the interviewees' statements.

\subsubsection{External Validity}
We have primarily focused on the automotive domain and the particular challenges originating from practical aspects and regulations therein. While some aspects may be insightful for other domains outside the automotive context, we did not investigate such aspects for this study. Furthermore, while many automotive OEMs are facing similar challenges in the development of safe ADS, deeper insights into their specific ways of working as well as methodological approaches are difficult as ways about \emph{how} to address rapid yet safe CDDM is considered a competitive edge. In our work, hence, we have focused on reporting challenges as the largest denominator for the automotive domain allowing us to generalize from our industrial context to competitors within the automotive domain. 


\section{Conclusion \& Future Work}
\label{sec:conlusion}

CDDM shall not be seen only as a challenge, but can also contribute towards a better engineering of safety-critical functions spanning from the development phases at OEMs into the actual operational phases when the vehicle is in customers' hands: It does not only increase the development rapidity of the software applications and their continuous feature expansion, but it is also an enabler for reacting to identified safety issues and possible design flaws by continuously receiving feedback from the field--which is a clear advantage for safety-critical functions to react swiftly. In this study, we have reported particular challenges for CDDM when applied to AD software development as reported by industrial practitioners. We conducted 8 interviews, aggregated the reported experience reports and opinions therefrom, and presented and discussed the identified challenges. We have complemented the presented challenges with relevant information from standards for the development of safety systems where applicable.

Future work needs to be conducted to explore further the opportunities from CDDM for AD and in particular, how it can be methodologically integrated into a safe engineering approach that starts already at the design phase to anchor the necessary product- and process-related aspects within an organization, but which also tightly integrates the monitoring of fleets of systems during their operations. Accomplishing that would allow to continuously monitor AD functions and their KPIs at scale within their ODDs, which enables early detection of potential issues or predict the performance of future function expansions.



\end{document}